\documentclass[%
 reprint,
nofootinbib,
 amsmath,amssymb,
 aps,
]{revtex4-2}

\usepackage{graphicx}
\usepackage{dcolumn}
\usepackage{bm}
\usepackage{bbm}
\usepackage[colorlinks=true, allcolors=blue]{hyperref}

\newcommand\barparena[1]{\overset{%
   \scriptscriptstyle(-)}{#1}}

\newcommand{\dd}{\mathrm{d}}
\newcommand{\ii}{\mathrm{i}}

\newcommand{\GF}{G_{\mathrm{F}}}
\newcommand{\abs}[1]{\left\lvert#1\right\rvert}
\newcommand{\norm}[1]{\left\lvert\left\lvert#1\right\rvert\right\rvert}

\begin{document}

\title{Spectral diversity in collisional neutrino-flavor conversion: flavor equipartition or swap}

\author{Masamichi Zaizen}
\affiliation{Department of Earth Science and Astronomy, The University of Tokyo, Tokyo 153-8902, Japan}

\date{\today}

\begin{abstract}
Quantum kinetics of neutrinos are known to potentially change the classical neutrino radiation field in high-energy astrophysical sources such as core-collapse supernovae and binary neutron-star mergers.
However, the mixing phenomena still have open issues in the nonlinear dynamics and the asymptotic states, particularly for recently discovered collision-induced flavor conversion.
In this paper, we investigate linear and nonlinear dynamics of collisional neutrino-flavor conversion (CFC) with multi-energy neutrino gases through numerical simulations, demonstrating that the asymptotic states dramatically change depending on unstable modes dominating the system.
In one unstable mode, high-energy neutrinos reach a flavor equipartition, but low-energy neutrinos return back to almost their initial states.
In contrast, in the other one, rather low-energy neutrinos achieve a full flavor swap, but high-energy neutrinos undergo less flavor conversion.
We clarify the distinct spectral behaviors in two different ways based on stability analysis and flavor pendulum.
Our result suggests that CFC with flavor swap can become crucial at deeper radii with low electron fraction and requires more detailed theoretical modeling of neutrino quantum kinetics.
\end{abstract}

\maketitle

\section{Introduction}
Theoretical modeling of neutrino flavor conversion has been conducted for the extreme environments such as core-collapse supernovae (CCSNe) and binary neutron star merger (BNSM) remnants \cite{Sigl:1993,Duan:2010,Tamborra:2021,Richers:2022b,Volpe:2024,Fischer:2024,Yamada:2024}.
The exploration has been motivated by discovering the occurrence region of flavor instability \cite{Dasgupta:2017,Abbar:2020,Nagakura:2021b,Akaho:2024a,Wu:2017,George:2020,Richers:2022a,Just:2022,Grohs:2024,Froustey:2024a}, which can dramatically change the neutrino radiation field in the classical transport scheme.
Also, with a phenomenological approach that assumes neutrino oscillations in a parametric way, it has been observed that many astrophysical properties are modified \cite{Li:2021,Just:2022,Fernandez:2022,Ehring:2023,Ehring:2023a,Ehring:2024,Nagakura:2024,Mori:2025}.
However, flavor conversion due to their self-interactions are principally nonlinear phenomena with much complexity, and the detailed physical process and the implementation is still under completion.

To accurately incorporate the effects of neutrino quantum kinetics into numerical modeling of the high-energy astrophysical events, the understanding of the asymptotic states is necessary.
Currently, two kinds of flavor instabilities have been attractively attracted: fast instability (FFI) \cite{Sawyer:2009} and collisional instability (CFI) \cite{Johns:2023a}.
Numerical simulations for fast flavor conversion triggered by FFI have been conducted in both local \cite{Bhattacharyya:2021,Bhattacharyya:2022,Wu:2021,Zaizen:2023,Zaizen:2023a,Xiong:2023b,George:2024,Nagakura:2025} and global geometry \cite{Nagakura:2022a,Nagakura:2023,Shalgar:2023,Xiong:2024}, and have revealed that it reaches the asymptotic state with flavor equipartition to eliminate angular crossings, which make the system unstable.
Note that advection via geometry and injection from collisions potentially brings further flavor conversion beyond a flavor equipartition \cite{Nagakura:2023b,Zaizen:2024,Fiorillo:2024b,Liu:2025}, and the potential and the implementation in CCSNe and BNSM remnants still need to be investigated.

On the other hand, the understanding of collisional flavor conversion (CFC) is still developing, except for CFI in the linear regime.
Collisions with matter generally decohere neutrino correlation between two flavors, and suppression or damping of FFI has been numerically observed 
\cite{Martin:2021,Shalgar:2021a,Kato:2021,Kato:2022,Sasaki:2022a,Hansen:2022,Johns:2022,Padilla-Gay:2022a,Fiorillo:2024a,DelfanAzari:2024}.
However, when the flavor-dependent decoherence is coupled with each other through neutrino self-interactions, they mutate into a flavor instability \cite{Johns:2023a}.
From linear stability analysis, CFI has two distinct regimes called resonance-like and otherwise, which are also quite different in the growth rates \cite{Xiong:2023c,Liu:2023}.
The growth rate of the standard CFI is dominated by collision rate $\propto R$ , whereas the resonance-like one is much faster and proportional to $\propto (\mu R)^{1/2}$, where $\mu$ denotes the self-interaction potential.
Both of them have been searched in the state-of-the-art simulations of CCSNe and BNSM remnants, and potentially emerge at deep radii where neutrinos are less transparent against matter \cite{Liu:2023c,Akaho:2024a,Xiong:2023c}.

Meanwhile, some notable features have been revealed in the nonlinear studies \cite{Lin:2023,Johns:2023c,Xiong:2023,Kato:2024}.
One of them is that the growth rate can be evaluated with energy-averaged flavor-decohering collision rates even in the multi-energy system, similar to in the monochromatic case \cite{Lin:2023}.
Other intriguing features are collisional flavor equipartition \cite{Lin:2023,Johns:2023c} and collisional flavor swap \cite{Kato:2024}.
Neutrinos with enough high reaction rates simultaneously undergo a flavor conversion and an irreversible flavor decoherence, leading to a complete depolarization, {\it i.e.} flavor equipartition.
Meanwhile, collisional swap occurs regardless of neutrino energies inside the resonance-like regime where the growth rate is fast enough to ignore the collision rates.
It should be noted that the resonance-like CFI leads to a flavor equipartition in the absence of the diagonal parts of the collision term, which classically changes neutrino populations and is often neglected in the quantum kinetic equation (QKE).

Such intriguing features of CFC have been discovered, but the dynamics of asymptotic behaviors, especially with multi-energy neutrinos, has been less explored.
Understanding this is necessary to develop the subgrid model of CFC and move it towards the more accurate theoretical modeling of CCSNe and BNSM remnants.
In this study, we demonstrate the spectral behaviors of CFC by employing some parametrized neutrino reaction rates.
The present paper also explores the dynamics by means of eigenvectors for unstable modes and flavor pendulum, providing descriptions for the asymptotic states establishing flavor equipartition or flavor swap and for the monotonicity over neutrino energy.

This paper is organized as follows.
In Sec.\,\ref{Sec:2_method}, we describe quantum kinetic equation leading collision-induced flavor conversion.
We demonstrate linear and nonlinear behaviros of CFC with two representative parameters for neutrino reaction rates in Sec.\,\ref{Sec:3_result}.
We then analyze the mechanism of the asymptotic states in two distinct ways in Sec.\,\ref{Sec:4_analysis}.
FInally, further discussions and conclusions will be given in Sec.\,\ref{Sec:5_conclusion}.

\section{Method}\label{Sec:2_method}
\subsection{Quantum kinetic equation}
QKE for dense neutrino gases is described by a neutrino density matrix $\rho$ as
\begin{align}
    \left(\partial_t +\boldsymbol{v}\cdot\nabla\right) \rho = -\ii \left[\mathcal{H}_{\nu\nu}, \rho\right] + \mathcal{C}[\rho],
\end{align}
where $\mathcal{C}$ denotes a collision term and $\mathcal{H}_{\nu\nu}$ a neutrino self-interaction Hamiltonian with neglecting vacuum and matter oscillations.
The self-interaction term is given by
\begin{equation}
    \mathcal{H}_{\nu\nu} = \sqrt{2}\GF \int\dd\Gamma_{\nu}^{\prime}\, v^{\mu}v_{\mu}^{\prime}\, \rho^{\prime},
\end{equation}
where the volume element in momentum space is defined by
\begin{equation}
    \int\dd\Gamma_{\nu} \equiv \int^{+\infty}_{-\infty}\frac{E_{\nu}^2\dd E_{\nu}}{2\pi^2}\int_{\Omega_{\nu}}\frac{\dd\Omega_{\nu}}{4\pi}.
\end{equation}
Here, the flavor-isospin convention, $\bar{\rho}(E_{\nu}) = -\rho(-E_{\nu})$, is employed as a negative energy occupation for the anti-neutrino density matrix.
In this paper, we investigate a flavor mixing by collision-induced flavor instability, particularly, focusing the spectral structures in multi-energy system.
For that aim, isotropic and homogeneous but multi-energy neutrino gases are assumed for the simplicity.
Then, multi-angle term in the self-interactions and spatial advection vanish in the QKE.
Also, we adopt the collision term only with the off-diagonal parts:
\begin{equation}
    \mathcal{C}[\rho] \equiv - R_{E}\, \rho_{ex},
\end{equation}
leading only to a flavor decoherence in the quantum regime and neglecting the classical effects relaxing flavor distributions in the Boltzmann neutrino transport.
Since we assume isotropic neutrino gases, reaction channels with scattering kernels are dropped in this way.

To perform the numerical simulations and detailed analyses, we employ the polarization vector configuration within a two-flavor framework:
\begin{equation}
    \rho = \frac{1}{2}P_0\mathbbm{1}_2 + \frac{f_{\nu_e}^{i}-f_{\nu_x}^{i}}{2}\boldsymbol{P}\cdot\boldsymbol{\sigma},
    \label{eq:Pauli}
\end{equation}
where $\boldsymbol{\sigma}$ denotes the Pauli matrices, $\boldsymbol{P}$ the polarization vector, $P_0 = \mathrm{Tr}\rho$, and $f_{\nu_{\alpha}}$ an occupation number distribution for a flavor $\alpha$.
In our notations, the polarization vector is normalized by the difference in the initial occupation number distributions between $\nu_e$ and $\nu_x$, and then the initial value of $P_3$, corresponding to the diagonal parts in the density matrix, is unity.
Within the configuration, the QKE is recast into 
\begin{equation}
    \dd_t \boldsymbol{P} = \boldsymbol{H}\times\boldsymbol{P} - R_E \boldsymbol{P}_{\perp},
    \label{eq:EoM_polari}
\end{equation}
where $\boldsymbol{P}_{\perp}$ is the transverse part of the polarization vector.
Now, we omit the classical collisional effects, i.e., the diagonal part of the matrix form of collisions, so that $P_0$ is conserved in our simulations.

\subsection{Linear stability analysis}
Linear stability analysis can be conducted by evaluating the off-diagonal components of the density matrix as perturbatives \cite{Izaguirre:2017}.
We restore the dimentions in phase space, and then the governing equation is
\begin{align}
    \ii\left(\partial_t +\boldsymbol{v}\cdot\nabla\right) \rho_{ex} &= \sqrt{2}\GF\, \rho_{ex} v^{\mu}\int\dd\Gamma_{\nu}^{\prime}\, v_{\mu}^{\prime} (\rho_{ee}^{\prime}-\rho_{xx}^{\prime}) \notag\\
        &\,\,\, -\sqrt{2}\GF\, (\rho_{ee}-\rho_{xx}) v^{\mu}\int\dd\Gamma_{\nu}^{\prime}\, v_{\mu}^{\prime} \,\rho_{ex}^{\prime} \notag\\
        &\,\,\, - \ii R_E \rho_{ex}.
\end{align}
With the plane-wave ansatz $\rho_{ex} \propto \tilde{Q}\exp(-\ii K^{\mu}x_{\mu})$, 
\begin{align}
    &\left[v^{\mu}(K_{\mu}-\Phi_{\mu})+\ii R_E\right]\tilde{Q} \notag\\
        &\,\,\,\,\,= -\sqrt{2}\GF\, (\rho_{ee}-\rho_{xx}) v^{\mu}\int\dd\Gamma_{\nu}^{\prime}\, v_{\mu}^{\prime} \,\tilde{Q}^{\prime},
        \label{eq:linear_Qeq}
\end{align}
where $\Phi_{\mu} = \sqrt{2}\GF\int\dd\Gamma\, v_{\mu}(\rho_{ee}-\rho_{xx})$.
Hereafter, we redefine a phase-shifted wave frequency $k_{\mu} \equiv K_{\mu}-\Phi_{\mu}$, and then the eigenvector satisfies the following equation:
\begin{equation}
    \tilde{Q} = (\rho_{ee}-\rho_{xx})\frac{v^{\mu}\tilde{a}_{\mu}}{v^{\sigma}k_{\sigma}+\ii R_E},
    \label{eq:eigen_Q}
\end{equation}
where
\begin{equation}
    \tilde{a}_{\mu} \equiv -\sqrt{2}\GF\,\int\dd\Gamma_{\nu}^{\prime}\, v_{\mu}^{\prime} \,\tilde{Q}^{\prime}.
\end{equation}
Inserting this equation into Eq.\,\eqref{eq:linear_Qeq} yields
\begin{equation}
    \Pi^{\mu\nu}\tilde{a}_{\nu} = 0
    \label{eq:Pi_a_eq}
\end{equation}
where
\begin{equation}
    \Pi^{\mu\nu}(k) = \eta^{\mu\nu} + \sqrt{2}\GF\int\dd\Gamma_{\nu}\, (\rho_{ee}-\rho_{xx})\frac{v^{\mu}v^{\nu}}{v^{\sigma}k_{\sigma}+\ii R_E}.
\end{equation}
The dispersion relation, which is a non-trivial solution for Eq.\,\eqref{eq:Pi_a_eq}, follows $\det\left[\Pi^{\mu\nu}(k)\right]=0$.
If there are imaginary parts in $k_{\mu}$, the flavor coherence $\rho_{ex}$ can exponentially grow or damp in space and time.

Now, we focus on collision-induced flavor instability in the isotropic and homogeneous neutrino background.
Under the environment, the self-interaction potential $\Phi_{j=1,2,3}$ bringing the phase shift to the wave number $\boldsymbol{k}$ is zero.
Thereby, a so-called zero mode $\boldsymbol{k}=0$ coincides with a ``true'' homogeneous mode $\boldsymbol{K}=0$.
The dispersion relation for the homogeneous mode is consequently given by
\begin{align}
    I \equiv \sqrt{2}\GF\int\frac{E_{\nu}^2\dd E_{\nu}}{2\pi^2} \frac{\rho_{ee}-\rho_{xx}}{\omega+\ii R_E} = -1 \,\,\mathrm{or}\,\, 3
    \label{eq:DR_CFI}
\end{align}
Complex wave frequency $\omega$ possibly appears when flavor-dependent collisional decoherence is included and leads to a collisional flavor instability \cite{Johns:2023a}.

Eq.\,\eqref{eq:DR_CFI} can be analytically solved for a monochromatic neutrino distribution \cite{Johns:2023a,Padilla-Gay:2022a}, and in the multi-energy case, the approximated solutions are provided with mean collision rates $\langle \barparena{R}\rangle$ in Refs.\,\cite{Lin:2023,Liu:2023}.
The approximated solutions are
\begin{equation}
    \omega_{\pm} = -A -\ii\gamma \pm \sqrt{A^2 - \alpha^2 + 2\ii G\alpha}
    \label{eq:CFI_sol}
\end{equation}
corresponding to an isotropy-preserving mode for $I=-1$ in Eq.\,\eqref{eq:DR_CFI}, which has larger growth rates compared than an isotropy-breaking mode for $I=3$ in the conditions \cite{Liu:2023}.
Each quantity is defined with (mean) collision rates and number densities:
\begin{equation}
    \gamma = \frac{R+\bar{R}}{2},\,\, \alpha = \frac{R-\bar{R}}{2},\,\, G = \frac{\mathfrak{g}+\bar{\mathfrak{g}}}{2},\,\, A = \frac{\mathfrak{g}-\bar{\mathfrak{g}}}{2},
    \label{eq:def_GA}
\end{equation}
with $\barparena{\mathfrak{g}}=\sqrt{2}\GF(\barparena{n}_{\nu_e}-\barparena{n}_{\nu_x})$.
In the following limits, Eq.\,\eqref{eq:CFI_sol} is simplified into easier fomulae \cite{Liu:2023,Akaho:2024a};
\begin{equation}
    \omega_{\pm} \approx 
    \begin{cases}
        -A -\ii\gamma \pm \left(\abs{A} +\ii\dfrac{G\alpha}{\abs{A}}\right)    \,\,\, &\mathrm{if}\, A^2 \gg \abs{G\alpha} \\
        -A -\ii\gamma \pm \sqrt{2\ii G\alpha}    \,\,\, &\mathrm{if}\, A^2 \ll \abs{G\alpha}.
    \end{cases}
    \label{eq:CFI_approx}
\end{equation}
The limit in the lower line is achieved only when $\abs{A}\sim 0$ and called a resonance-like CFI \cite{Xiong:2023c,Liu:2023,Kato:2024}.
In this work, we focus only on the corresponding limit to the upper case, and then the maximum growth rate is given by
\begin{equation}
    \mathrm{max}\left[\mathrm{Im}\,\omega_{\pm}\right] \approx 
        -\gamma +\abs{ \dfrac{G\alpha}{A}} \,\,\, \mathrm{if}\, A^2 \gg \abs{G\alpha},
    \label{eq:CFI_sol_apprx}
\end{equation}
which is roughly proportional to the collision rates $R$.

\subsection{Model}
Our aims are to understand the asymptotic behaviors of CFC.
To this end, we employ parametric neutrino energy distributions and flavor-dependent collision rates.
We assume initial neutrino systems are composed only of pure electron-type neutrinos for simplicity, though previous studies have revealed the contribution from heavy-leptonic flavors stabilizes the system in the CFI case \cite{Liu:2023c}.
Following Ref.\,\cite{Lin:2023}, initial neutrino Fermi-Dirac distribution is
\begin{equation}
    f_{\nu} \propto \frac{1}{\exp[(E_{\nu}-\mu_{\nu})/T_{\nu}]+1}
\end{equation}
with $\mu_{\nu}$ be a chemical potential set by zero and choosing $T_{\nu_e}=4\,\mathrm{MeV}$ and $T_{\bar{\nu}_e}=5\,\mathrm{MeV}$.
And instead of the chemical potentials which arrange the hierarchy of neutrino number densities, the asymmetric parameter of $\bar{\nu}_e$ to $\nu_e$ is set $\alpha_{\mathrm{asym}}\equiv n_{\bar{\nu}_e}/n_{\nu_e} = 0.8$.
And we set the self-interaction strength is $\mu_0 = 10^4\,\mathrm{km^{-1}}$ and an initial perturbation of $10^{-3}$ in the off-diagonal component instead of the vacuum term.
Energy dependence of collision rates is for simplicity adopted by
\begin{equation}
    \barparena{R}_{\nu}(E_{\nu}) = \barparena{R}_0\left(\frac{E_{\nu}}{10\,\mathrm{MeV}}\right)^2,
\end{equation}
where we fix $R_0=1\,\mathrm{km^{-1}}$ for neutrinos and parametrize $\bar{R}_0$ for anti-neutrinos from $0$ to $2\,\mathrm{km^{-1}}$.

\begin{figure}[t]
    \centering
    \includegraphics[width=1.\linewidth]{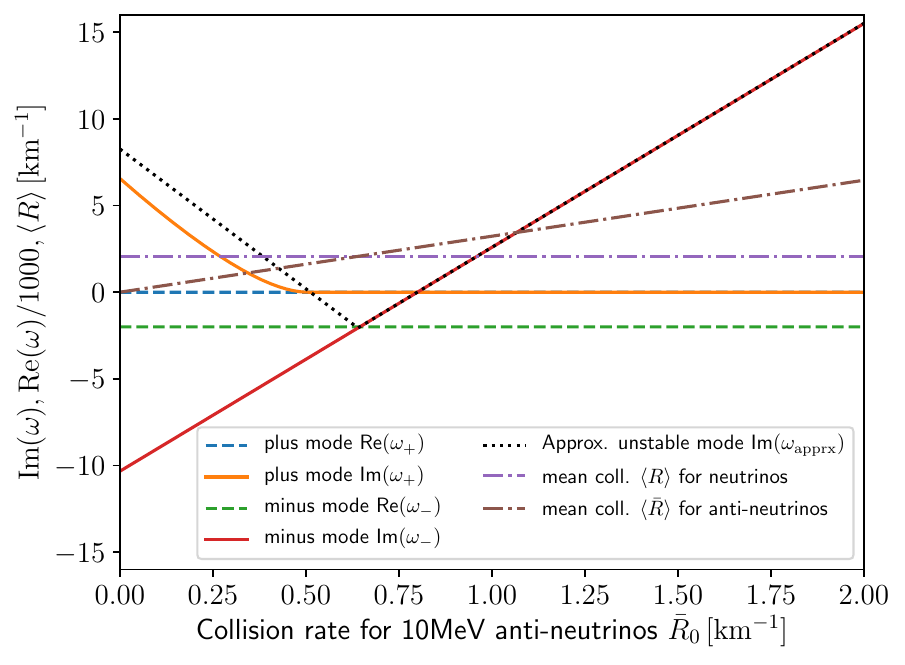}
    \caption{Collisional instability modes by linear stability analysis as a function of reaction rates for antineutrinos.
    Two solid and dashed lines are imaginary and real parts of wave frequencies numerically solved from Eq.\,\eqref{eq:DR_CFI}, respectively.
    Dotted lines correspond to approximated scheme in Eq.\,\eqref{eq:CFI_sol_apprx}.
    Dot-dashed lines to mean collision rates for neutrinos and antineutrinos.}
    \label{fig:instability}
\end{figure}
Figure\,\ref{fig:instability} demonstrates collisional modes in flavor instability obtained numerically from Eq.\,\eqref{eq:DR_CFI} and solved analytically as Eq.\,\eqref{eq:CFI_sol_apprx} as a function of reaction rate $\bar{R}_0$ for antineutrinos.
It is clearly found that dominant growing modes switch around $\langle\bar{R}_E\rangle \simeq \langle R_E\rangle$ with the reaction rate increasing.
As can be seen from the definition in Eqs.\,\eqref{eq:def_GA} and \eqref{eq:CFI_approx}, the sign of $\alpha$ changes after that equality, and the maximum growing mode changes so as to compensate for that sign.
Note that we fix neutrino number densities in this study, but if not so, the change of the sign of $A$ can bring similar behaviors \cite{Johns:2023c}.
The disparity in the sign between $A$ and $\alpha$ produces the distinction between the plus and minus modes.
In the subsequent section, the two unstable modes display quite different behaviors in both linear and nonlinear dynamics.

\section{Multi-Energy Flavor Evolution}\label{Sec:3_result}
\begin{figure}[t]
    \centering
    \includegraphics[width=1.\linewidth]{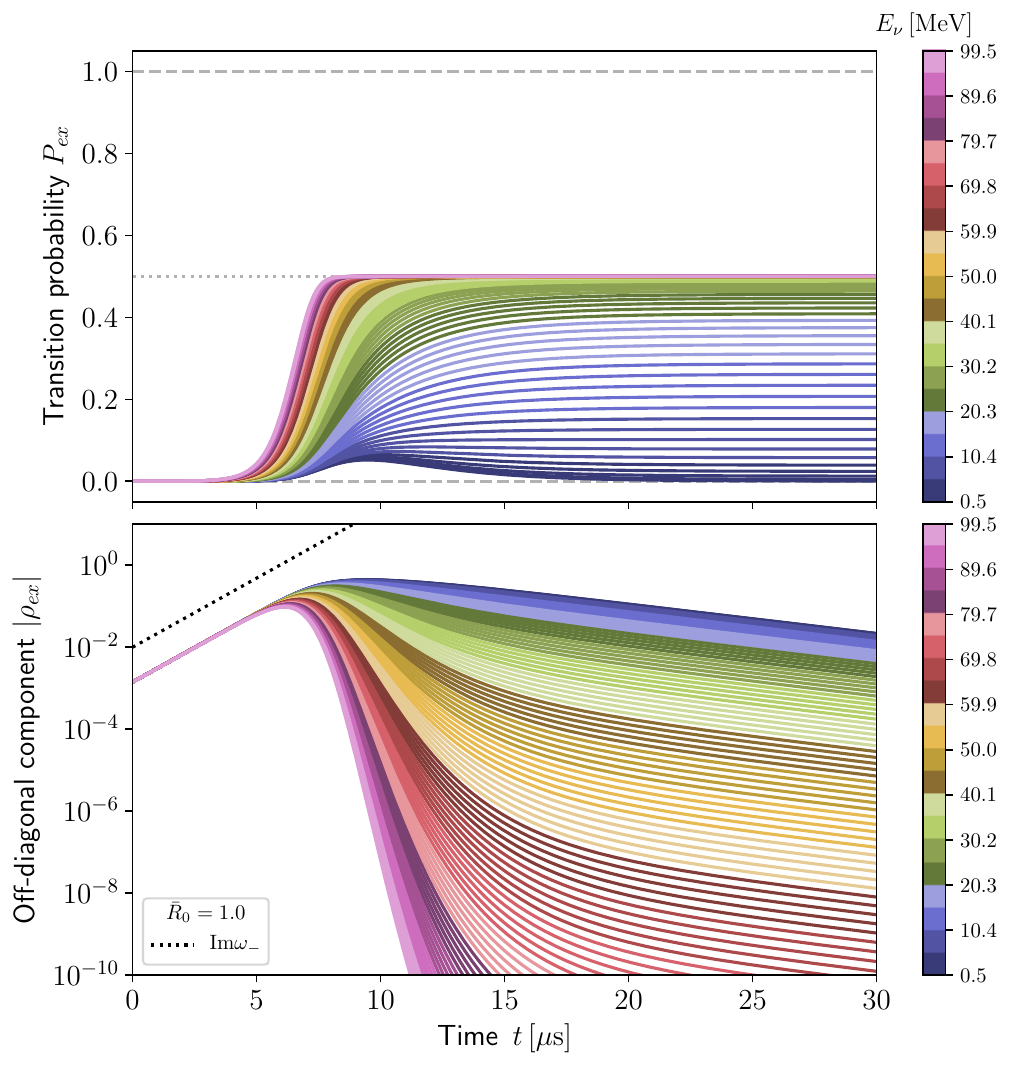}
    \caption{Time evolution of transition probability $P_{ex}$ (top) of electron-type neutrinos and flavor coherency $\abs{\rho_{ex}}$ (bottom), respectively.
    Collisional rates for antineutrinos are $\bar{R}_0 = 1\mathrm{\,km^{-1}}$, and color contours for lines correspond to neutrino energy dependence.
    Dotted line corresponds to unstable mode exhibited in Fig.\,\ref{fig:instability}.
    }
    \label{fig:prob_R1}
\end{figure}
We perform the two representative cases, adopting the higher and lower sides of reaction rates for antineutrinos, $\bar{R}_0 = 1$ and $0.1 \mathrm{\,km^{-1}}$.
These choices correspond to two unstable modes of CFI, the plus and minus modes, in Fig.\,\ref{fig:instability}.
The disparity in the reaction rates between neutrinos and antineutrinos can be attributed to the properties of background matter, such as electron fraction $Y_e$.
Actually, $\nu_e$ opacity is about an order of magnitude more dominant than that of $\bar{\nu}_e$ at the radii with $Y_e\sim0.1$, while in the $Y_e\sim0.5$ regions they are comparable \cite{Akaho:2024a}.

Figure\,\ref{fig:prob_R1} shows the time evolution of transition probability $P_{ex}(E_\nu)$ (top panel) of electron-type neutrinos and flavor coherence $\abs{\rho_{ex}(E_\nu)}$ (bottom) with energy dependency in the case of reaction rate $\bar{R}_0 = 1\mathrm{\,km^{-1}}$.
Neutrinos with the highest energy reach a flavor equipartition, $P_{ex}\sim 0.5$, at the asymptotic states, while lower ones have monotonically weaker transition probabilities in the top panel.
In the bottom, isoenergetic exponential growth proceeds in the linear phase, and then energy-dependent collisional decoherence makes the system return to the flavor eigenstates in the later phase.
This result is consistent with what was reported in the previous work \cite{Lin:2023}.
Note that in this figure, the flavor coherence $\abs{\rho_{ex}}$ or $\norm{\boldsymbol{P}_{\perp}}$ is normalized so as to make the initial $P_3$ unity.
This means that the spectral structure of the transition probability and the flavor coherence does not include the neutrino distribution functions.
Now, since we assume the same reaction rate for antineutrinos as that for neutrinos, such conversion properties become similar, and we do not present them in the figure.

\begin{figure}[t]
    \centering
    \includegraphics[width=1.\linewidth]{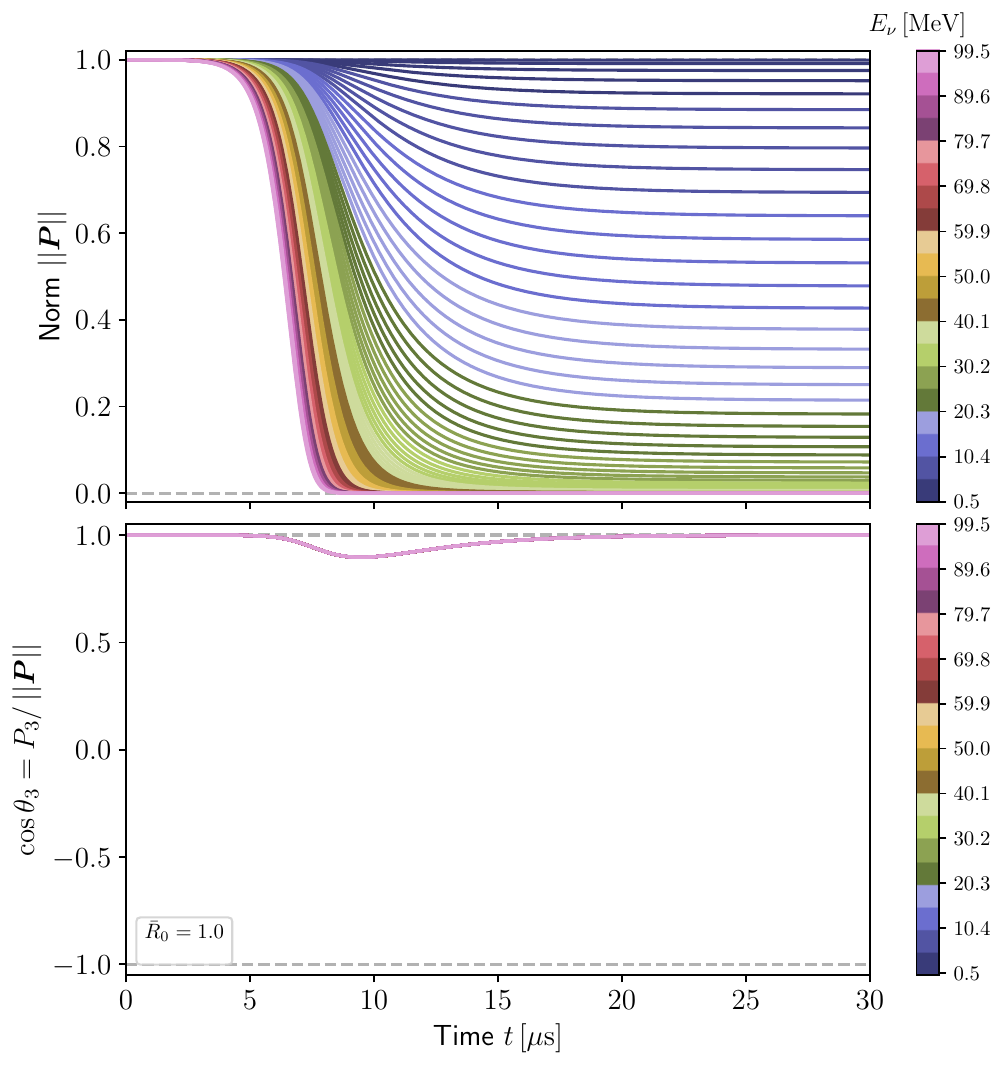}
    \caption{Time evolution of the norm $\norm{\boldsymbol{P}}$ (top) and the angle $\theta_3$ (bottom) relative to the $z$-axis of the polarization vector $\boldsymbol{P}$ in the flavor space for neutrinos.
    In the bottom panel, all colored lines are overlapped.}
    \label{fig:norm_R1}
\end{figure}
To cultivate the understanding of the nonlinear behaviors, we exhibit the time evolution of the norm of the polarization vector for neutrinos in the top panel of Fig.\,\ref{fig:norm_R1}.
The polarization vector with the highest neutrino energy completely collapses within the Bloch sphere, while the ones for neutrinos with lower energy remain almost at pure states.
The following equation gives this norm evolution;
\begin{equation}
    \dd_t \norm{\boldsymbol{P}}^2 = - 2 R_E \norm{\boldsymbol{P}_{\perp}}^2.
    \label{eq:norm_eq}
\end{equation}
Since the flavor coherence $\norm{\boldsymbol{P_{\perp}}}$ grows isoenergetically, the spectral structure of the norm shrinking is determined by the energy dependence of reaction rates $R_E$.
So, the length of the polarization vector shortens monotonically with $\propto E_{\nu}^2$ in the energy domain.
Meanwhile, the bottom panel of Fig.\,\ref{fig:norm_R1} exhibits the angle of the polarization vector relative to the third axis (hereafter denoted as the $z$-axis) in flavor space.
It clearly shows that the tilt of the polarization vector is independent of neutrino energy and the asymptotic state for each neutrino is aligned upwardly along the $z$-axis.
It means that leaning and standing the polarization vector are carried out by collisional decoherence shared through the self-interaction term.
Actually, neutrinos with the lowest energy receive only the contribution from the self-interaction Hamiltonian $H_{\nu\nu}$, which explicitly does not depend on the neutrino spectra, because the collision rate is extremely small.
Therefore, the asymptotic state is determined by the degree of collisional decoherence which they have experienced.

\begin{figure*}[t]
    \centering
    \includegraphics[width=1.\linewidth]{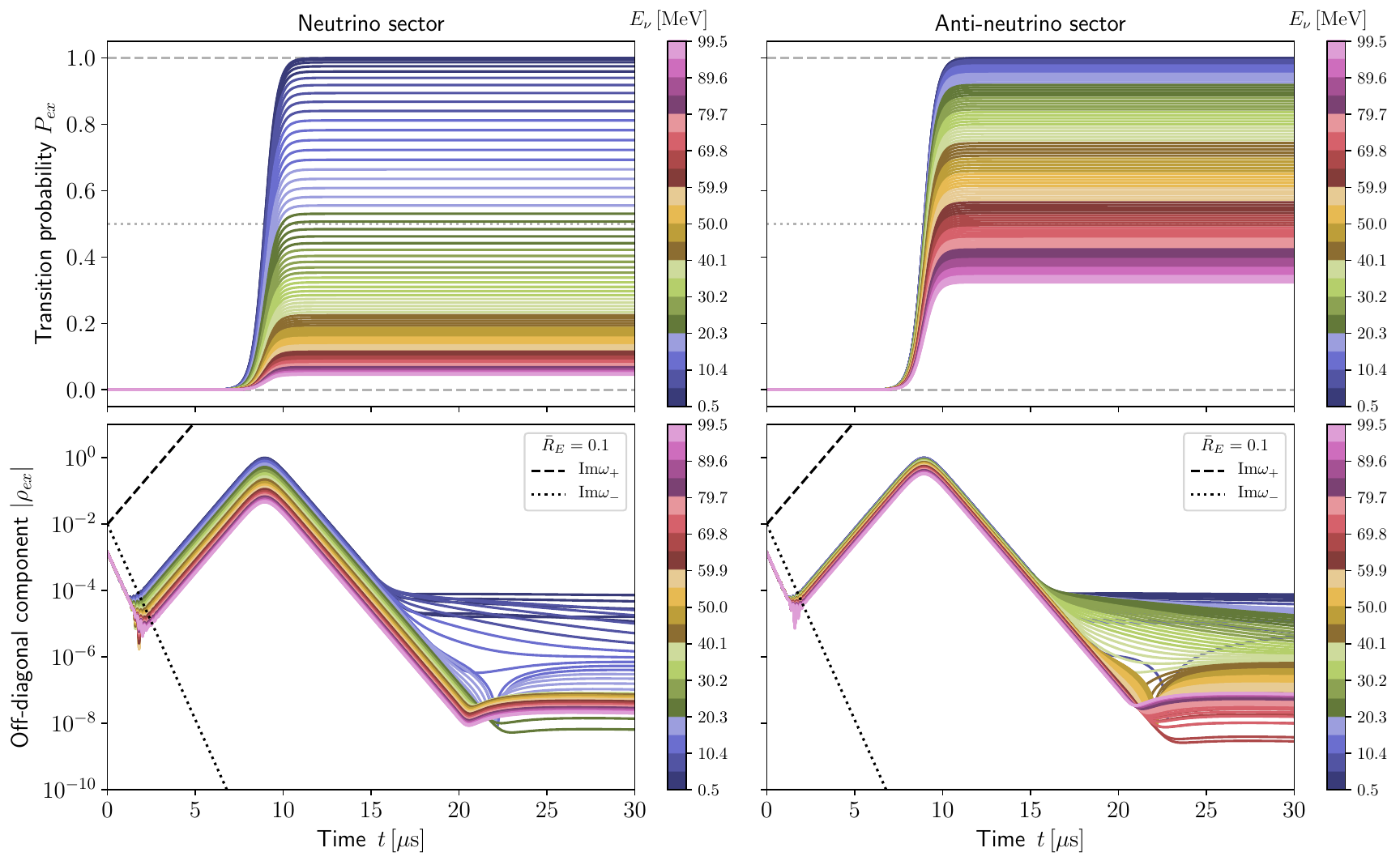}
    \caption{Same as Fig.\,\ref{fig:prob_R1}, but collision rates $\bar{R}_0=0.1\mathrm{\,km^{-1}}$ for antineutrinos.}
    \label{fig:prob_R01}
\end{figure*}
In the polarization vector configuration, the transition probability is given by
\begin{equation}
    P_{ex} = \frac{1}{2}(P_0 - P_3)/P_0.
\end{equation}
The asymptotic states are now settled down in flavor eigenstates because the flavor coherence decays through collisional decoherence.
Consequently, the length of the polarization vector is almost identical to $P_3$ and determines the transition probability as follows:
\begin{equation}
    P_{ex} = \frac{1}{2}(1 - \norm{\boldsymbol{P}}).
\end{equation}
Complete collapse of the polarization vector means that only the trace part $P_0$ of density matrix survives in Eq.\,\eqref{eq:Pauli} as resulting from collisional decoherence.
Since the contribution of heavy-leptonic flavor in the trace part is identical to that of electron-type one, the resultant flavor equipartition is established in the system.

When considering the quantum contribution of collision term, we can regard the entire system as the composite system composed of neutrinos and background matter\footnote{
Note that each subsystem is now assumed to be within the mean-field framework, so we do not take into account the entanglement among neutrinos themselves. See Ref.\,\cite{Balantekin:2023} on the detail about many-body corrections for self-interactions}.
Then, neutrinos, which are initially at pure states, undergo flavor conversion through linear growth but simultaneously are entangled with the surrounding matter via collisional decoherence.
The occurrence of entanglement can be generally expressed by that the Bloch vector for density matrix enters inside the Bloch sphere and is equivalent to the shrink of the polarization vector\footnote{Entropy as a measure of entanglement can be defined using the norm of the polarization vector, and the complete collapse of the polarization vector means that the entanglement entropy is maximized.}.
Hence, neutrinos that have been collisionally decohered can not return to the initial states because the density matrices differ entirely from those of neutrinos at pure states.
On the contrary, neutrinos less coupled to matter potentially remain at pure states.

This is the reason why neutrinos with low-energy undergo less shrink of the polarization vector and can go back to the initial state, $P_{ex}\sim 0$.
On the other hand, higher-energy neutrinos are exposed to inevitable collisional decoherence with growing flavor coherence, leading to a complete depolarization, corresponding to flavor equipartition, $P_{ex}\sim 0.5$.
Just after reaching a linear saturation, the system undergoes the competition between flavor conversion by nonlinear term and decoherence effect by collision term.
Since the collisional decoherence wastes the matured flavor coherence and damps the system, the entire system consequently returns back to the linear order, and the collapsed norm determines the asymptotic states.
Thereby, the tilt of the polarization vector is so tiny, as can be seen in Fig.\,\ref{fig:norm_R1}.

\begin{figure}
    \centering
    \includegraphics[width=1.\linewidth]{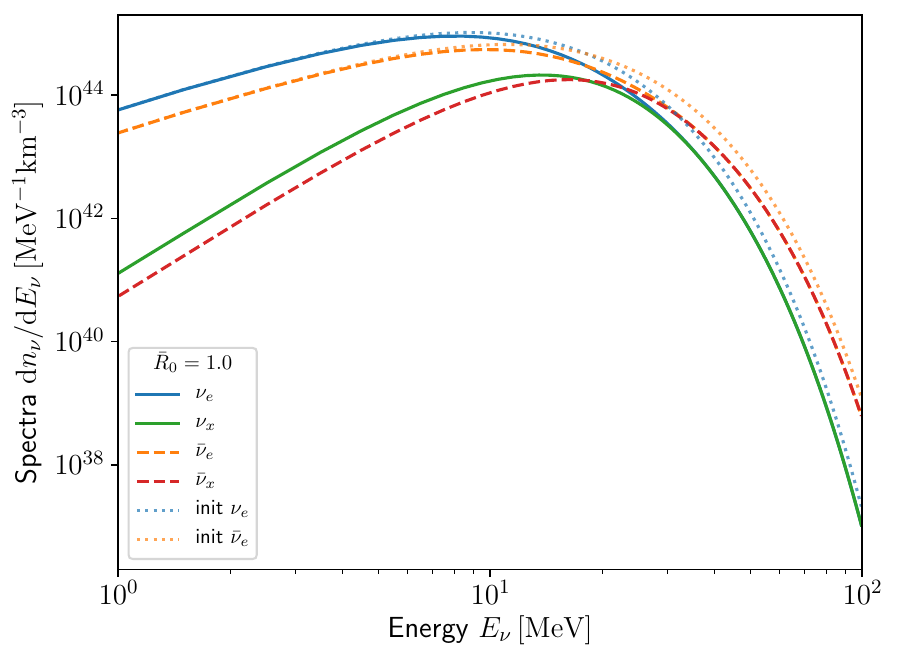}
    \includegraphics[width=1.\linewidth]{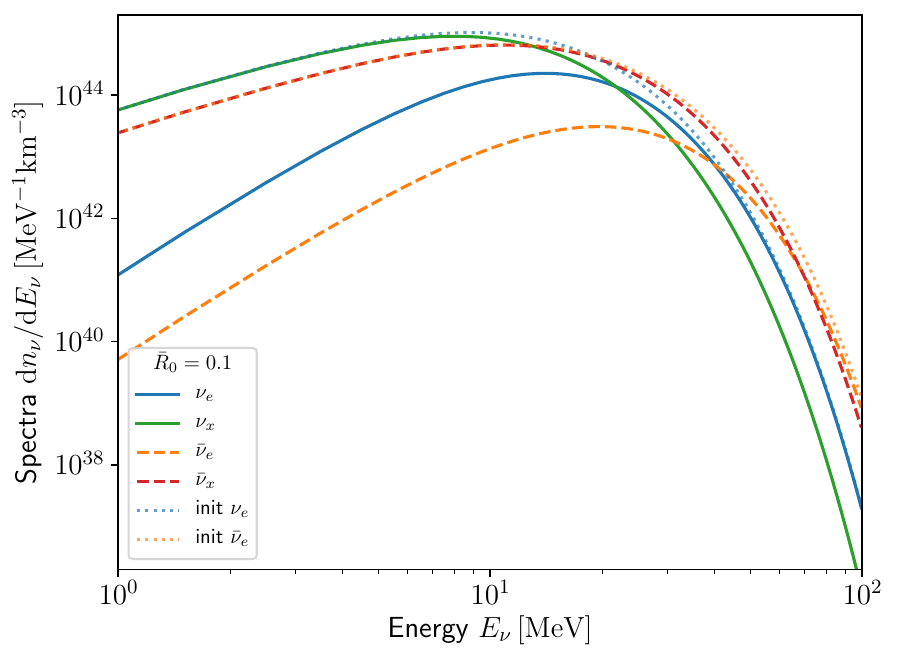}
    \caption{
    Energy distribution in the case of $\bar{R}_0=1$ (top) and $\bar{R}_0=0.1\,\mathrm{km^{-1}}$ (bottom).
    Dotted lines are initial spectra for electron-type neutrinos.
    Solid and dashed ones correspond to final one for neutrinos and antineutrinos after CFC, respectively.
    Clearly, the order of colored lines ($\barparena{\nu}_e$ and $\barparena{\nu}_x$) is exchanged on the low-energy side between the two scenarios due to the presence or absence of collisional swap.
    }
    \label{fig:spectra}
\end{figure}
\begin{figure*}[t]
    \centering
    \includegraphics[width=1.\linewidth]{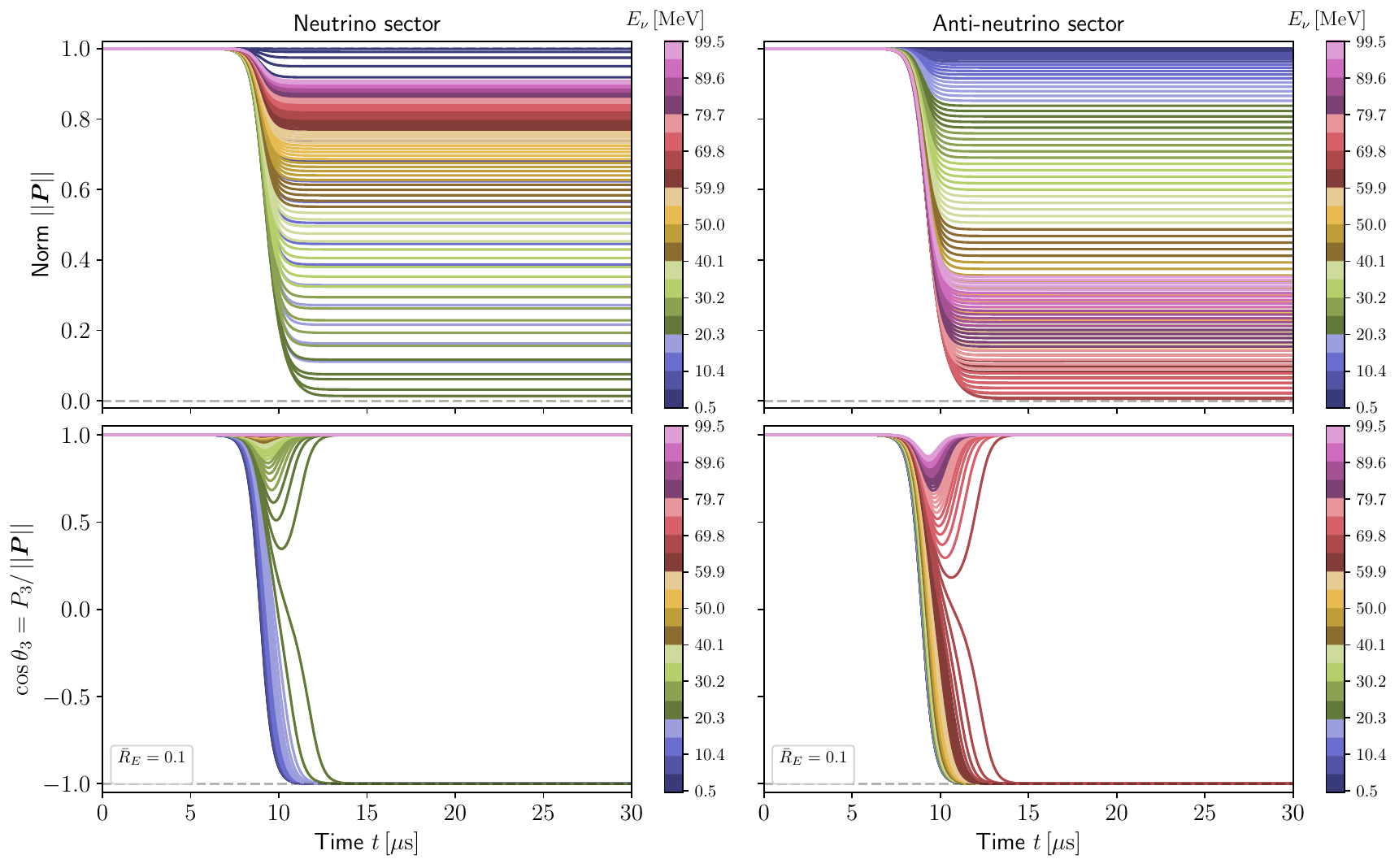}
    \caption{Same as Fig.\,\ref{fig:norm_R1}, but plots in the both neutrino and antineutrino sectors for $\bar{R}_0=0.1\mathrm{\,km^{-1}}$.}
    \label{fig:angle_R01}
\end{figure*}
However, linear and nonlinear behaviors significantly change when we employ lower antineutrino reaction rates.
Figure\,\ref{fig:prob_R01} demonstrates the flavor evolution in the case of $\bar{R}_0 = 0.1\mathrm{\,km^{-1}}$.
We find that neutrinos with the lowest energy establish a complete flavor swap, $P_{ex}=1$, while the higher energy one rather remains close to electron-type flavor, $P_{ex}<0.5$.
This result is quite contrary to the case of $\bar{R}_0 = 1\mathrm{\,km^{-1}}$, and means that a flavor swap can occur even in the absence of the diagonal parts of collision term unlike the previous works \cite{Kato:2024} in the case with a resonance-like CFI.
Also, in the bottom panel, the flavor coherence first undergoes linear damping isoenergetically, and then exponential growth with some energy spreading occurs.
The behaviors can be understood through the unstable modes in the linear regime.
As clarified in Fig.\,\ref{fig:instability}, this simulation setup has two types of non-stable modes, minus $\omega_-$ and plus modes $\omega_+$.
As the antineutrino reaction rate decreases, the imaginary part $\mathrm{Im}(\omega_-)$ of the minus mode reduces and becomes negative.
On the other hand, that $\mathrm{Im}(\omega_+)$ of the plus mode increases and switches into the growing mode.
This means that the distinction between the two simulations above stems from the difference in the dominant growing modes.
Comparing the energy distributions at the asymptotic states also clarifies the distinction as presented in Fig.\,\ref{fig:spectra}.
In the low-energy side, the order of final spectra of $\barparena{\nu}_e$ and $\barparena{\nu}_x$ are exchanged because collisional swap occurs in the case of $\bar{R}_0 = 0.1\,\mathrm{km^{-1}}$ (bottom).
Also, in the case of $\bar{R}_0 = 1\,\mathrm{km^{-1}}$ (top), the complete overlap between two flavors appears in the high-energy side through a flavor equipartition.

Figure\,\ref{fig:angle_R01} shows the time evolution of the norm $\norm{\boldsymbol{P}}$ and the angle $\cos\theta_3$ relative to the $z$-axis of the polarization vector in the case of $\bar{R}_0 = 0.1\mathrm{\,km^{-1}}$.
Unlike the case of $\bar{R}_0 = 1\mathrm{\,km^{-1}}$, the monotonicity in the spectral distribution of the norm is broken.
Also, the bottom panel displays that the polarization vectors appear to be spectrally split.
But the reason for such distinct behaviors can be clarified by Eq.\,\eqref{eq:norm_eq}.
In this calculation setup, the flavor coherence already achieves some spectral structures even in the linear phase in Fig.\,\ref{fig:prob_R01}.
Therefore, the evolution of the norm depends on both reaction rates and flavor coherency over neutrino energy.
The characteristics for each unstable mode can be read from the eigenvector $\tilde{Q}_E$ in the subsequent sections.

\section{Flavor Analysis}\label{Sec:4_analysis}
\subsection{Flavor Analysis by Eigenvector}
Under the assumptions of isotropic and homogeneous neutrino background, the eigenvector associated with the unstable mode in Eq.\,\eqref{eq:eigen_Q} can be recast into 
\begin{equation}
    \tilde{Q}_E = (\rho_{ee}-\rho_{xx})\frac{1}{\omega + \ii R_E}
\end{equation}
and the amplitude is given by
\begin{equation}
    \abs{\tilde{Q}_E} = (\rho_{ee}-\rho_{xx})\frac{1}{\sqrt{\omega_r^2 + (\omega_i+R_E)^2}},
    \label{eq:amp_Qv}
\end{equation}
where $\omega = \omega_r + \ii \omega_i$.
The energy dependence can be produced by reaction rates, except for the neutrino distributions.
Recalling Eq.\,\eqref{eq:CFI_approx}, the real parts of $\omega_{\pm}$ are approximated to $\mathrm{Re}(\omega_{\pm}) = -A \pm \abs{A}$.
When the sign of $A$ is positive, the real part of the plus mode vanishes, while the minus mode has $-2A$ in the real part.
In the limit of $A^2 \gg \abs{G\alpha}$, the value of $A$ (or the self-interaction potential) is much larger than reaction rates.
This means that the amplitude of the eigenvector is $\abs{\tilde{Q}_E}\propto \abs{\omega_r}^{-1}$ in the minus mode and becomes independent of neutrino energy because $\omega_r$ exceeds the other terms in the denominator in Eq.\,\eqref{eq:amp_Qv}.
On the other hand, the plus mode $\omega_+$ becomes almost pure complex, and so the amplitude has energy dependence with $(\omega_i+R_E)^{-1}$.

\begin{figure}[t]
    \centering
    \includegraphics[width=1.\linewidth]{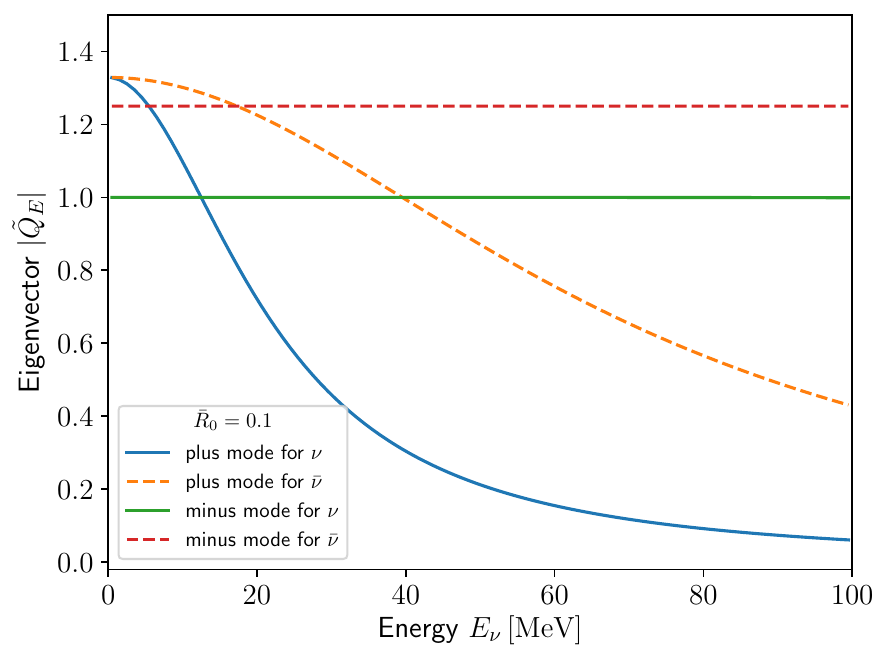}
    \caption{Amplitude of eigenvector $\tilde{Q}_E$ for plus mode $\omega_+$ and minus mode $\omega_-$ as a function of neutrino energy for $\bar{R}_0 = 0.1\mathrm{\,km^{-1}}$.
    Solid lines are for neutrinos and dashed lines for antineutrinos.
    The amplitude is normalized with the number density.}
    \label{fig:amp_Qv}
\end{figure}
Figure\,\ref{fig:amp_Qv} demonstrates the energy dependence of the eigenvector $\tilde{Q}_E$ for both the plus and minus modes in the case of $\bar{R}_0 = 0.1\mathrm{\,km^{-1}}$.
As noted above, the minus mode is homogeneous in neutrino energy, and the plus mode has a distribution peaked at lower energy.
Note that the eigenvector for antineutrinos becomes flatter because the collision rate $\bar{R}_E$ is weaker than neutrinos.
The eigenvector says that flavor coherency should be evolved with the energy distribution associated with the corresponding unstable mode.
This fact is consistent with the isoenergetic exponential growth in Fig.\,\ref{fig:prob_R1} for $\bar{R}_0 = 1\mathrm{\,km^{-1}}$ because the plus mode is stable and only the minus mode is active in this setup.
On the other hand, in the case of $\bar{R}_0 = 0.1\mathrm{\,km^{-1}}$, both the minus and plus modes have non-zero imaginary parts.
Also, in our simulation setup, the initial seed of flavor coherence is homogeneously given in neutrino energy.
Hence, the minus mode with the homogeneous eigenvector, which is a damping mode, reduces the flavor coherence at first, and then the spectral component with $\propto (\omega_i+R_E)^{-1}$  following the plus mode appears excited.
As confirmed from the bottom panel of Fig.\,\ref{fig:prob_R01}, neutrinos with the lower energy begin to grow before the higher-energy neutrinos.
Also, in the antineutrino sector (right panel), the spectral deviation is smaller than in the neutrino sector (left) because the spectral structure of the eigenvector is less dependent on neutrino energy.
Note that when we employ an initial seed from the vacuum term, not the artificial perturbation, the damping phase in the early epoch does not appear much because the perturbation has a spectral structure with a vacuum frequency $\omega_V \propto E_{\nu}^{-2}$.

The difference in the nonlinear behaviors between Figs.\,\ref{fig:prob_R1} and \ref{fig:prob_R01} can be understood by the spectral structure in the linear growth phase.
In the case of $\bar{R}_0=1\mathrm{\,km^{-1}}$, all neutrinos with any energy simultaneously reach a linear saturation, and then collisional decoherence becomes to dominate the system evolution, as can be seen in Fig.\,\ref{fig:prob_R1}.
On the other hand, in the case of $\bar{R}_0=0.1\mathrm{\,km^{-1}}$, since the eigenvector of growing mode has a low-energy peaked distribution, neutrinos with lower energy reach the linear saturation before the higher energetic ones.
The nonlinear power propagates into the other neutrinos via their self-interactions, and neutrinos with higher energy are forced to behave as in the nonlinear regime.
However, the higher energetic neutrinos themselves do not have enough significant flavor coherence $\rho_{ex}$ yet, so the resultant collisional decoherence is still weak.
As a consequence, collisional decoherence in the entire system can not prevent the low-energy neutrinos from undergoing a continuous flavor conversion in the nonlinear order, and they establish a complete flavor swap.

\begin{figure}[t]
    \centering
        \includegraphics[width=1.\linewidth]{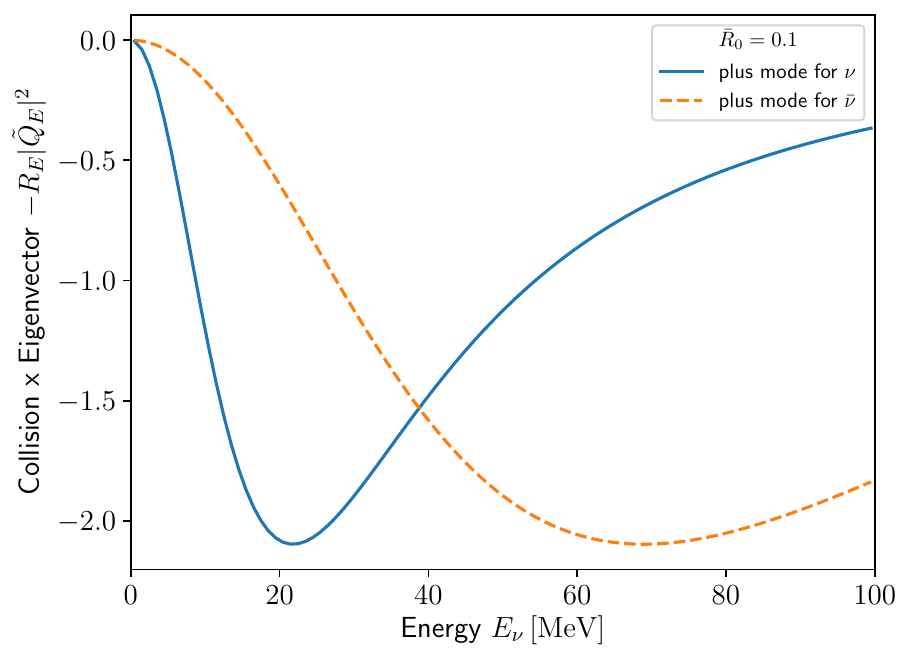}
    \caption{Energy distribution of the multiplications of collisions and squared norm, corresponding to the r.h.s. of Eq.\,\eqref{eq:norm_eq}.}
    \label{fig:col_Qv}
\end{figure}
The balance between flavor coherence and collisional decoherence can be read out through the energy dependence of the norm collapse presented in Fig.\,\ref{fig:col_Qv}.
The distribution has the extremum in the middle of the considered neutrino energy range because the collision rate is proportional to $E_{\nu}^2$ whereas the flavor coherency is peaked at low energy.
Actually, in the top panel of Fig.\,\,\ref{fig:angle_R01}, neutrinos (thick green) with $\sim20\mathrm{\,MeV}$ and antineutrinos (red) with $\sim 70\mathrm{\,MeV}$ are completely collapsed within the Bloch sphere.
The fully depolarized (anti-)neutrinos establish a flavor equipartition as shown in Fig.\,\ref{fig:prob_R01}.
On the other hand, since neutrinos with the lowest energy continue to hold the length of the polarization vector, they can reach a full flavor swap.
Also, neutrinos with the highest energy can not completely get away from shrinking the length of the polarization vectors, so they are at mixed states, which are not at pure states only with electron-type flavors.

The extremum energy point separates whether neutrinos go across the $x-y$ plane in flavor space or a flavor equipartition line $P_3=0$, as seen in Fig.\,\ref{fig:angle_R01}.
From the above discussion about the flavor swap, this extremum is an indicator because it corresponds to the neutrino energy where the collision term maximumly affects the flavor conversion.
In the lower energy side of Fig.\,\ref{fig:col_Qv}, the reaction rates are too weak to brake the nonlinear flavor conversion due to matured flavor coherence, and so the system leads to a full or partial flavor swap beyond the equipartition line.
On the higher energy side, the flavor coherence is too small to drive a large flavor conversion and also to collapse the state but with strong collisions, so the system goes back near the initial state.
At the extremum energy point, both effects are enough strong to lead to a flavor equipartition.
The suggestion coincides with the polarization vectors appearing to be monotonically split at the critical energy in Fig.\,\ref{fig:angle_R01}.

\subsection{Flavor Analysis by Pendulum Model}
With a flavor pendulum, we can also understand why the polarization vectors for the low-energy neutrinos can rotate to the opposite orientation.
To this end, we consider the equation of motion for the flavor pendulum, the same as Refs.\,\cite{Johns:2023c,Kato:2024}.
Unlike slow and fast flavor pendula, CFI can be described as the energy exchange between flavor pendulum and the environment.
Unlike their ways, our calculations require multi-energy approaches, which are not integrated.
We begin with Eq.\,\eqref{eq:EoM_polari}.
We introduce the sum and difference vectors $\boldsymbol{S}_E\equiv \boldsymbol{P}_E+\bar{\boldsymbol{P}}_E$ and $\boldsymbol{D}_E\equiv \boldsymbol{P}_E-\bar{\boldsymbol{P}}_E$.
Then, the equation of motion is given by
\begin{align}
    \dot{\boldsymbol{S}}_E &= \mu\boldsymbol{D}^{\mathrm{int}}\times \boldsymbol{S}_E - R^{+}_E \boldsymbol{S}_{\perp,E} - R^{-}_E \boldsymbol{D}_{\perp,E} \\
    \dot{\boldsymbol{D}}_E &= \mu\boldsymbol{D}^{\mathrm{int}}\times \boldsymbol{D}_E - R^{-}_E \boldsymbol{S}_{\perp,E} - R^{+}_E \boldsymbol{D}_{\perp,E},
\end{align}
where $R^{\pm}_E \equiv (R_E\pm\bar{R}_E)/2$ and $\boldsymbol{D}^{\mathrm{int}}$ denotes the self-interaction potential, which is $\boldsymbol{D}_E$ integrated over neutrino energy.
Note that since we now take into account the energy dependency, the term $\mu\boldsymbol{D}^{\mathrm{int}}\times \boldsymbol{D}_E$ still survives.

We want to know the nonlinear behaviors of the polarization vector with the lower energy.
So, we focus on the extremely low-energy neutrinos, $E_{\nu} \sim 0$.
Since collision rates are proportional to $E_{\nu}^2$, the extreme components are insensitive to the background matter.
Hence, the equation of motion can be recast by dropping the collision term into 
\begin{align}
    \dot{\boldsymbol{S}}_{E\sim0} &= \mu\boldsymbol{D}^{\mathrm{int}}\times \boldsymbol{S}_{E\sim0} \label{eq:Sdot} \\
    \dot{\boldsymbol{D}}_{E\sim0} &= \mu\boldsymbol{D}^{\mathrm{int}}\times \boldsymbol{D}_{E\sim0},
\end{align}
while the time derivative of the energy-integrated $\boldsymbol{D}^{\mathrm{int}}$ is
\begin{align}
    \dot{\boldsymbol{D}}^{\mathrm{int}} &=  - \langle R^{-}_E \boldsymbol{S}_{\perp,E}\rangle - \langle R^{+}_E \boldsymbol{D}_{\perp,E}\rangle \notag \\
        &\approx - \langle R^{-}_E \rangle  \boldsymbol{S}^{\mathrm{int}}_{\perp} - \langle R^{+}_E\rangle \boldsymbol{D}^{\mathrm{int}}_{\perp} \notag \\
        &\approx - \langle R^{-}_E \rangle  \boldsymbol{S}^{\mathrm{int}}_{\perp}.
        \label{eq:Dint_eq}
\end{align}
We here approximate $\langle R_E \boldsymbol{S}_E\rangle \approx \langle R_E\rangle \boldsymbol{S}^{\mathrm{int}}$ from the first line to the second and $\norm{\boldsymbol{S}^{\mathrm{int}}}\gtrsim\norm{\boldsymbol{D}^{\mathrm{int}}}$ from the second to the third.
Then, the second derivative of the sum $\boldsymbol{S}_E$ is 
\begin{align}
    \ddot{\boldsymbol{S}}_{E\sim0} &= \mu\dot{\boldsymbol{D}}^{\mathrm{int}}\times \boldsymbol{S}_{E\sim0} + \mu\boldsymbol{D}^{\mathrm{int}}\times \dot{\boldsymbol{S}}_{E\sim0} \notag \\
        &\approx -\mu \langle R^{-}_E \rangle  \boldsymbol{S}^{\mathrm{int}}_{\perp} \times\boldsymbol{S}_{E\sim0} + \mu\boldsymbol{D}^{\mathrm{int}}\times \left(\mu\boldsymbol{D}^{\mathrm{int}}\times \dot{\boldsymbol{S}}_{E\sim0}\right) \notag \\
        &\approx \mu^2\left(\boldsymbol{D}^{\mathrm{int}}\cdot\boldsymbol{S}_{E\sim0}\right)\boldsymbol{D}^{\mathrm{int}} - \mu^2 \norm{\boldsymbol{D}^{\mathrm{int}}}^2 \boldsymbol{S}_{E\sim0},
\end{align}
where we adopt the condition, $\mu\gg R_E$, from the second line to the thrid.
The acceleration of the polarization vector along the $z$-axis is given by the following:
\begin{equation}
    \ddot{S}_{z,E\sim0} \approx \mu^2\left[\boldsymbol{D}^{\mathrm{int}}\cdot\boldsymbol{S}_{E\sim0}\right] D^{\mathrm{int}}_{z} - \mu^2 \norm{\boldsymbol{D}^{\mathrm{int}}}^2 S_{z,E\sim0}.
    \label{eq:S_ddot}
\end{equation}
This equation is the same as Eq.\,(24) in Ref.\,\cite{Kato:2024}, which describes collisional flavor swap under the resonance-like CFI.
Within this regime, the growth rate is fast enough to ignore the contribution from the collision term.
The extremely low-energy limit we now assume is identical to the situation in the previous work.
However, the criterion of whether neutrinos can reach a complete swap or go beyond the flavor equipartition is different.
If the acceleration to the negative direction is continually negative, the polarization vector can be dropped to the opposite $z$-direction beyond the flavor equipartition plane.
The sign of the second term is determined by $S_{z,E\sim0}$, which is the quantity we want to know.
It is initially positive and becomes close to zero when it approaches the flavor equipartition plane.
And, when $S_{z}\lesssim 0$, the second term becomes positive.
If the first term is also positive at that time, the acceleration to the positive direction activates and would oscillate around the flavor equipartition plane.
In that case, the polarization vector can never go to $S_{z} \sim -1$.

To achieve a flavor swap, the first term needs to be held negative.
Now, since classical collisions, which vary neutrino populations in the diagonal components, are neglected in our setup, $D^{\mathrm{int}}_z$ is constant in time, from Eq.\,\eqref{eq:Dint_eq}.
Therefore, the sign of the first term is determined by the term $\left[\boldsymbol{D}^{\mathrm{int}}\cdot\boldsymbol{S}_{E\sim0}\right]$.
The time derivative is from Eqs.\,\eqref{eq:Sdot} and \eqref{eq:Dint_eq}
\begin{equation}
    \dd_t\left[\boldsymbol{D}^{\mathrm{int}}\cdot\boldsymbol{S}_{E\sim0}\right] \approx - \langle R^{-}_E \rangle  \boldsymbol{S}^{\mathrm{int}}_{\perp} \cdot \boldsymbol{S}_{E\sim0}.
    \label{eq:Dint_SE}
\end{equation}
This sign depends mainly on the difference $\langle R^{-}_E \rangle$ in the mean collision rates, corresponding to energy-dependent $\alpha$ in Eq.\,\eqref{eq:def_GA}.
As can be seen from Fig.\,\ref{fig:instability}, the sign changes with employed collision rates for antineutrinos, and at the same time, the dominant growing modes also switch from the plus into the minus ones.
In other words, the sign of the derivative determines which growing modes dominate the system.
For the minus mode, $\langle R^{-}_E \rangle$ is negative and Eq.\,\eqref{eq:Dint_SE} becomes positive, so that Eq.\,\eqref{eq:S_ddot} potentially becomes positive.
Consequently, the polarization vector is prohibited to promote flavor conversion beyond the flavor equipartition line.
On the other hand, for the plus mode, $\langle R^{-}_E \rangle$ is positive and Eq.\,\eqref{eq:Dint_SE} becomes negative, so that the acceleration in Eq.\,\eqref{eq:S_ddot} can continue to be negative.
Thereby, the polarization vector with lower energy or weaker reaction rates can achieve a flavor swap beyond the flavor equipartition line.
This suggestion is consistent with the fact that the energy range where the polarization vector can go across the equipartition line is wider in the antineutrino sector compared to in the neutrino sector in Fig.\,\ref{fig:angle_R01} because the limitation where the collision term is small enough to be neglected can be applied for the broader energy range.
\begin{figure}
    \centering
    \includegraphics[width=1.\linewidth]{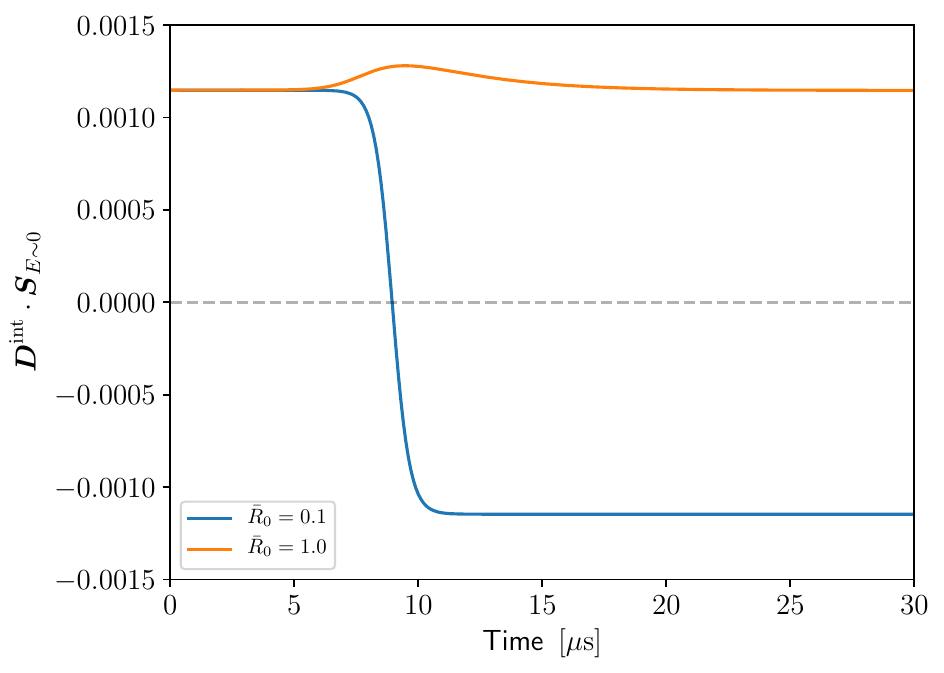}
    \caption{Time evolution of $\boldsymbol{D}^{\mathrm{int}}\cdot\boldsymbol{S}_{E\sim0}$ in Eqs.\,\eqref{eq:S_ddot} and \eqref{eq:Dint_SE} for both cases of $\bar{R}_0 = 0.1$ and $\bar{R}_0 = 1\mathrm{\,km^{-1}}$.}
    \label{fig:Dint_SE}
\end{figure}
The numerical evolution of the term $\left[\boldsymbol{D}^{\mathrm{int}}\cdot\boldsymbol{S}_{E\sim0}\right]$ is exhibited in Fig.\,\ref{fig:Dint_SE}, and the case of $\bar{R}_0 = 0.1\mathrm{\,km^{-1}}$ clearly presents a decrease to negative while the other case is always positive.
The result coincides with the above discussions on the sign of the derivative and the behaviors of the sign around the flavor equipartition line.
Based on the equations and the demonstrations, it is found that whether neutrinos weakly coupled with matter lead to a flavor equipartition or flavor swap is determined by the magnitude of mean collision rates between neutrinos and antineutrinos, that is, the dominant growing modes.

\section{Summary and Conclusions}\label{Sec:5_conclusion}
In this paper, we have presented multi-energy behaviors in linear and nonlinear regimes of collisional neutrino-flavor conversion (CFC) in homogeneous and isotropic backgrounds.
There are two unstable modes in collisional flavor instabilities (CFI) triggering CFC, and whether or not one of them prominently evolves the system in the linear phase is determined by the magnitude relation in both mean collision rates and number densities between neutrinos and antineutrinos.
We have demonstrated that CFC has spectral diversity in the asymptotic states, from no conversion to complete flavor swap or equipartition.
We have investigated what spectral structures are yielded in the asymptotic behaviors and how the associated flavor instabilities drive the dynamics in terms of linear stability analysis and flavor pendulum.

In the case where the number density of neutrinos exceeds that of antineutrinos, a flavor instability, called a plus mode, emerges when the mean collision rates have the same magnitude relation, while the so-called minus mode becomes excited when the relation is twisted.
Our numerical simulations showed that the minus mode leads the system to a flavor equipartition, but rather, the plus mode favors a flavor swap.
Unlike the behaviors reported in Ref.\,\cite{Kato:2024}, this flavor swap can occur with energy dependence even in the inclusion of only flavor-decohering collision term and even outside the resonance-like region of CFI.
The monotonicity of CFC in neutrino energy was found to be the opposite between the two unstable modes.
Lower-energy neutrinos go back to the initial flavor eigenstates in the case of the minus mode, while they rather promote flavor conversion across the flavor equipartition in the case of the plus mode.
The opposite flavor evolution can be understood in two different ways: linear stability analysis and flavor pendulum.

In the linear regime, the unstable mode evolves with the plane wave corresponding to the eigenvector, which can be obtained from the dispersion relation.
We have analytically obtained the associated eigenvector using the approximated scheme for the growth rate presented in Ref.\,\cite{Liu:2023}.
Numerical solutions have confirmed our analytic formulae that the plus mode has a spectral structure with a peak at the low energy side, while the minus mode is homogeneous in neutrino energy.
The linear behavior is consistent with the nonlinear simulations based on the quantum kinetic equations and can describe the subsequent nonlinear dynamics in CFC.

In the case of the plus mode, the lower energy components reach a linear saturation before the higher ones and drive the nonlinear behavior into the system.
Since they are less coupled with background matter, only the self-interaction term promotes flavor conversion without collisional decoherence.
At the same time, the higher-energy neutrinos feel the nonlinear power via the self-interactions even though they construct less flavor coherence.
Thereby, they do not undergo collisional decoherence much and can not stop the flavor conversion in the entire system.
The resultant asymptotic states settle down at a full or partial flavor swap in the components with weaker reaction rates.
On the other hand, in the case of the minus mode, the eigenvector is homogeneous in neutrino energy, and all neutrinos reach a linear saturation simultaneously.
The isoenergetic evolution maximizes the impact of collisional decoherence, and the system is immediately pulled back from nonlinear order to linear.
Consequently, the asymptotic states are determined only by the norm of the polarization vector for the neutrino density matrix, which shrinks due to the collisional decoherence.

Far from the qualitative descriptions of flavor evolution, the nonlinear behaviors can also be explained with a flavor pendulum.
In the extremely low-energy limit $E_{\nu}\sim0$, which corresponds to the extremely weak reaction rate limit $R_E\sim0$, the motion of the flavor pendulum can be simplified only with the self-interactions.
We have found that the criterion of whether flavor conversion can proceed beyond the flavor equipartition line is given by the difference in mean collision rates between neutrinos and antineutrinos.
The criterion is the same as whether the plus or minus modes drive the system in the linear regime.
Thereby, the suggestion coincides with the fact that neutrinos weakly coupled with matter can achieve a flavor swap through the self-interaction term in the case of the plus mode.

Although we have presented various descriptions for the dynamics of CFC in multi-energy framework, there remains some crucial issues.
First, we have neglected the contribution from the diagonal components in the collision term, which drives classical neutrino transport.
The inclusion is required to take into account the feedback on the background matter in CCSNe and BNSMs, and it can also modify the asymptotic behaviors of CFC because it changes the length of the polarization vector.
Actually, in the case with a resonance-like CFI on which we have not focused in this work, the presence of collisional flavor swap, which was first reported in Ref.\,\cite{Kato:2024}, requires including the neutrino-population changing parts.
Second, we have assumed isotropic neutrino distributions throughout our study.
It has been revealed that CFI can emerge at deeper radii, where the baryon density is roughly between $10^{10}$ and $10^{12}\mathrm{\,g\,cm^{-3}}$, within CCSNe compared than FFI in Refs.\,\cite{Liu:2023c,Akaho:2024a}, and this is the rationale that we can drop the angular distributions of neutrinos in our simulations.
However, neutrinos can achieve forward-peaked distributions outside flavor-dependent decoupling radii, and the radii can enter the occurrence region of CFI.
In that situation, the isotropic assumption is broken, and the scattering process with matter can be efficient.
On the other hand, the stability analysis for scattering has not yet been explored, and the impact has been less studied.
There are still many studies to be needed, but our presenting insights in this paper will lead the future works.

\section{Acknowledgments}
We thank Hiroki Nagakura, Lucas Johns, and Yudai Suwa for valuable comments and discussions.
This work is supported by Grant-in-Aid for JSPS Fellows (Grant No. 22KJ2906) and JSPS KAKENHI Grant Numbers JP24H02245 and JP25K17383.

\bibliography{papers}

\end{document}